\documentclass[pdflatex,sn-mathphys-num]{sn-jnl}

\usepackage{graphicx}
\usepackage{multirow}%
\usepackage{amsmath,amssymb,amsfonts}%
\usepackage{amsthm}%
\usepackage{mathrsfs}%
\usepackage[title]{appendix}%
\usepackage{xcolor}%
\usepackage{textcomp}%
\usepackage{manyfoot}%
\usepackage{booktabs}%
\usepackage{rotating}%
\usepackage{algorithm}%
\usepackage{algorithmicx}%
\usepackage{algpseudocode}%
\usepackage{listings}%
\usepackage{hyperref}
\usepackage{tabularx}
\usepackage{array}
\usepackage{listings}

\begin{document}

\title[MagPot]{MagPot: A magnetoelastic potential generator for molecular dynamics simulations}

\author*[1]{\fnm{P.} \sur{Nieves}}\email{nievespablo@uniovi.es}
\author[2]{\fnm{I.} \sur{Korniienko}}
\author[2]{\fnm{S.} \sur{Arapan}}
\author[3]{\fnm{A.} \sur{Fraile}}
\author[1]{\fnm{R.} \sur{Iglesias}}
\author[2]{\fnm{D.} \sur{Legut}}

\affil*[1]{\orgdiv{Departamento de F\'{i}sica}, \orgname{Universidad de Oviedo}, \orgaddress{\street{C. Leopoldo Calvo Sotelo, 18}, \city{Oviedo}, \postcode{33007}, \state{Asturias}, \country{Spain}}}

\affil[2]{\orgdiv{IT4Innovations}, \orgname{V\v{S}B - Technical University of Ostrava}, \orgaddress{\street{17. listopadu 2172/15}, \postcode{70800}, \state{Ostrava}, \country{Czech Republic}}}
\affil[3]{\orgname{Instituto Catalán de Nanociencia y Nanotecnología (ICN2)},\orgaddress{\postcode{08193}, \city{Barcelona}, \country{Spain}}}

\abstract{Magnetoelastic phenomena originate from the coupling between atomic displacements and magnetic degrees of freedom, and their atomistic simulation requires interatomic models in which magnetic interactions depend explicitly on interatomic distance. This paper presents \textsc{MagPot}, a Python program that generates Bethe–Slater parameter sets for spin–lattice molecular dynamics simulations using the SPIN package of the Large-scale Atomic/Molecular Massively Parallel Simulator (LAMMPS). From a compact set of material properties, the program determines the radial functions describing Heisenberg exchange and the dipole and quadrupole contributions of the two-ion Néel interaction. Version 1.0 supports simple-cubic (SC), body-centred-cubic (BCC) and face-centred-cubic (FCC) lattices with ferromagnetic (FM) and antiferromagnetic (AFM) configurations. A central feature of \textsc{MagPot} is its modular architecture, which allows the magnetic contribution it generates to be combined additively with a broad range of compatible non-magnetic interatomic potentials. The parameterisation workflow is assessed through equation-of-state, Curie-temperature, volume-magnetostriction, pressure-dependent magnetocrystalline-anisotropy and anisotropic-magnetoelastic tests. By automating the construction of the magnetic contribution to
the interatomic model, \textsc{MagPot} offers a straightforward and reproducible route from material properties to LAMMPS-ready
magnetic interactions, facilitating large-scale atomistic
simulations of magnetoelasticity and magnetostriction, as well as physical phenomena associated with magnon-phonon coupling.}

\maketitle

\section*{Article Highlights}
\begin{itemize}
  \item \textit{\textsc{MagPot} converts material properties into magnetic interaction parameters ready for LAMMPS simulations.}
  \item \textit{Its modular output can be combined with a wide range of different types of potentials.}
  \item \textit{Version 1.0 covers three cubic lattices and selected ferromagnetic and antiferromagnetic orders.}
\end{itemize}

\section{Introduction}
\label{sec:introduction}

Magnetoelasticity describes the coupling between the magnetic and mechanical degrees of freedom of a solid. In a magnetic material, changes in the atomic positions modify the interactions between magnetic moments, while changes in the magnetic state can induce lattice deformations. This mutual coupling is at the origin of a broad range of phenomena, including magnetostriction, magnetically induced changes in the elastic response, magnetovolume effects and the coupling between magnons and phonons \cite{lacheisserie_magnetostriction_1993,engdahl_handbook_1999}. These effects are relevant both to fundamental studies of magnetic materials and to applications involving magnetic sensors, actuators, transducers and energy conversion \cite{dapino_encyclopedia_2000}. They can also have a substantial influence on thermomechanical properties, for example through changes in thermal expansion and phonon transport near magnetic phase transitions \cite{wasserman_chapter_1990}.

Conventional molecular dynamics (MD) describes the motion of the atoms but generally does not include the dynamical evolution of their magnetic moments. Conversely, atomistic spin dynamics describes the magnetic degrees of freedom while the atomic positions are usually fixed or their motion is neglected \cite{eriksson_atomistic_2017}. These separate descriptions are insufficient for studying magnetoelastic phenomena in which atomic positions and magnetic moments are strongly coupled. Spin--lattice dynamics combines molecular dynamics and atomistic spin dynamics, allowing the atomic positions and magnetic moments to evolve simultaneously \cite{tranchida_massively_2018}. In this framework, the dependence of the magnetic interactions on interatomic distance provides a direct mechanism for coupling the two subsystems. Such an approach is particularly useful for investigating magnetoelastic and magnetocaloric processes over length and time scales that are difficult to access using first-principles molecular dynamics alone.

A practical spin--lattice model must describe not only the non-magnetic interatomic interactions, but also the magnetic interactions and their dependence on the atomic configuration. For cubic crystals, Nieves \textit{et al.} developed a parametrization based on the Heisenberg exchange interaction and the two-ion Néel model \cite{nieves_spin-lattice_2021}. In that formulation, the distance dependence of the magnetic interactions is represented by Bethe--Slater functions. The exchange function $J(r)$ controls the isotropic magnetic interaction and can be parametrized in relation to the magnetic transition temperature and the spontaneous volume magnetostriction. The dipolar and quadrupolar Néel functions, $l(r)$ and $q(r)$, describe the magnetoelastic and magnetocrystalline-anisotropy contributions associated with spin--orbit coupling. Analytical relations were derived to connect the values and first derivatives of these functions at the equilibrium neighbour distance to material properties, including the magnetoelastic constants and the magnetocrystalline anisotropy constant. The procedure was originally demonstrated for cubic ferromagnetic systems \cite{nieves_spin-lattice_2021}, and more recently has been extended to antiferromagnets  \cite{korniienko_NiO_2026,Nieves_FeRh_2026}. The modularity of this approach has also been demonstrated in previous spin-lattice studies employing different reference potentials, including modified embedded atom method (MEAM) \cite{nieves_spin-lattice_2021}, Buckingham \cite{korniienko_NiO_2026}, Lennard--Jones \cite{Nieves_FeRh_2026}, reference-free
MEAM \cite{korniienko_NiO_2026}, and spectral neighbor analysis potential (SNAP) \cite{nikolov_temperature_2023}.

Although these analytical relations provide a systematic route for constructing spin--lattice models, applying them requires selecting the appropriate expressions for the crystal structure and magnetic order, handling the unit conversions, and translating the resulting parameters into the input format required by a simulation package. The purpose of the present work is to facilitate this procedure through \textsc{MagPot}, a Python-based generator of magnetic interaction parameters for spin--lattice simulations with the SPIN package of LAMMPS \cite{tranchida_massively_2018,thompson_lammps_2022,plimpton_fast_1995}. Starting from a compact input file containing structural, elastic and magnetic properties, \textsc{MagPot} evaluates the required values and radial derivatives of $J(r)$, $l(r)$ and $q(r)$, converts them into Bethe--Slater parameters, and writes the corresponding LAMMPS commands. The first release focuses on simple-cubic (SC), body-centred-cubic (BCC) and face-centred-cubic (FCC) lattices, considering the ferromagnetic and  antiferromagnetic configurations.

\textsc{MagPot} is complementary to the \textsc{MAELAS} software suite \cite{nieves_maelas_2021}. \textsc{MAELAS} was developed to automate first-principles calculations of magnetocrystalline anisotropy, anisotropic magnetostriction and magnetoelastic constants, using density functional theory (DFT) calculations \cite{nieves_maelas_2021,nieves_maelas_2022}. Subsequent developments extended the methodology to the calculation of exchange magnetostriction and related isotropic magnetoelastic properties \cite{nieves_automated_2023}. These capabilities make \textsc{MAELAS} a possible source of several material-specific quantities required to construct a spin--lattice model, such as $K_1$, its pressure dependence, and the anisotropic magnetoelastic constants. Other required input quantities, including the equilibrium distance and bulk modulus, can be obtained from an equation-of-state calculation, while magnetic transition temperatures may be taken from experiment or first-principles calculations. \textsc{MagPot} uses these quantities to construct the distance-dependent interactions required by the atomistic model.

\begin{figure}[htbp]
\centering
\includegraphics[width=\linewidth]{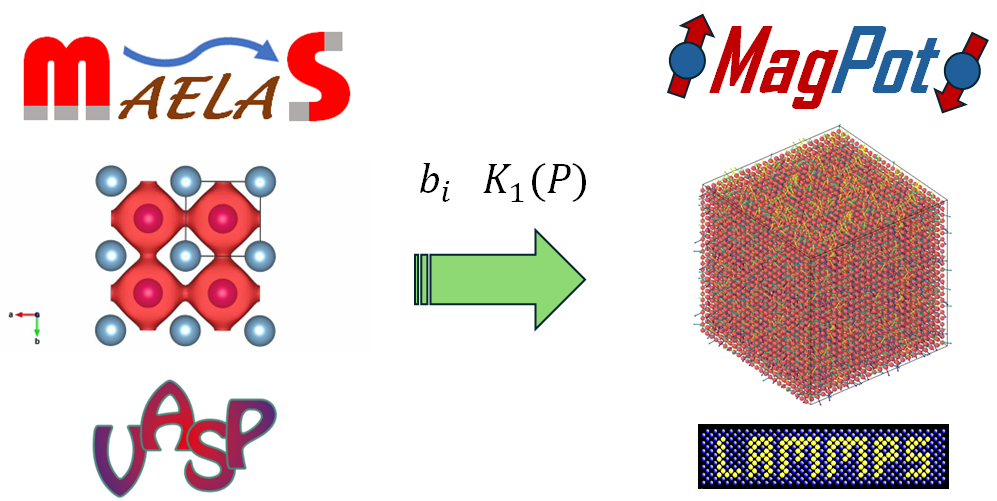}
\caption{Connection between the \textsc{MAELAS} and \textsc{MagPot} codes. \textsc{MAELAS} is interfaced with first-principles calculations codes like the Vienna Ab initio Simulation Package (VASP) \cite{kresse_efficiency_1996,kresse_ultrasoft_1999} to compute magnetoelastic constants ($b_i$) and the pressure dependence of magnetocrystalline anisotropy ($K_1(P)$) which are used as input parameters for \textsc{MagPot} to generate the  magnetoelastic potential suitable for large-scale spin-lattice molecular dynamics simulations with the software LAMMPS \cite{thompson_lammps_2022,tranchida_massively_2018}. }
\label{fig:maelas_magpot}
\end{figure}

The combination of first-principles calculations interfaced with \textsc{MAELAS} and \textsc{MagPot} therefore provides a possible multiscale workflow, see Fig.\ref{fig:maelas_magpot}. First-principles calculations supply material-specific magnetic and magnetoelastic information; \textsc{MagPot} translates this information into the parameters of the magnetic Hamiltonian in order to generate the required magnetoelastic potential to perform spin--lattice molecular dynamics in substantially larger systems and over longer times than are generally accessible to direct first-principles molecular dynamics. In this sense, the two software tools address complementary stages of the modelling process: \textsc{MAELAS} helps determine the material properties, whereas \textsc{MagPot} constructs the magnetic interaction functions used in the atomistic simulations.

The remainder of this paper is organised as follows. Section~\ref{sec:methods} describes the implementation of \textsc{MagPot}, the magnetic Hamiltonian and the workflow used to obtain its input parameters. Section~\ref{sec:examples} presents representative examples and tests of the generated interactions against the target magnetic and magnetoelastic properties. Finally, Section~\ref{sec:conclusions} summarises the capabilities of the current version and discusses possible future developments.

\section{Methods}
\label{sec:methods}

\subsection{Program overview, input and output}
\label{subsec:magpot_program}

\textsc{MagPot} is a lightweight program written in Python~3 that generates the magnetic pair-interaction coefficients required to perform spin--lattice molecular dynamics simulations with the SPIN package of LAMMPS \cite{thompson_lammps_2022,tranchida_massively_2018}. The source files of the program \textsc{MagPot} are available in the GitHub repository \url{https://github.com/pnieves2019/magpot}. Version~1.0 uses NumPy to read the input data and perform the numerical operations. The program does not generate the spin-independent reference potential; this contribution must be supplied separately (e.g., using embedded-atom
method (EAM) or MEAM potentials, pairwise models such as Lennard--Jones or Buckingham, or suitable machine-learning interatomic potentials (MLIPs) such as SNAP) and combined additively with the magnetic interactions in LAMMPS through the \texttt{hybrid/overlay} pair style.

The calculation in \textsc{MagPot} is controlled by a plain-text file named \texttt{input.dat}. Lines beginning with the character \texttt{\#} are comments and are ignored when the file is read. The remaining entries must be supplied in the prescribed order. The first three entries select the output format, crystal structure and magnetic order; the following eight entries are numerical material parameters. Table~\ref{tab:input_parameters} summarises the input variables and their units.

\begin{table}[htbp]
\centering
\caption{Input parameters read by \textsc{MagPot} version~1.0. The entries must appear in the order shown in the \texttt{input.dat} file.}
\label{tab:input_parameters}
\small
\setlength{\tabcolsep}{3.5pt}
\renewcommand{\arraystretch}{1.15}
\begin{tabularx}{\textwidth}{@{}>{\raggedright\arraybackslash}p{2.15cm}X>{\raggedright\arraybackslash}p{2.15cm}@{}}
\hline
\textbf{Entry} & \textbf{Description} & \textbf{Unit / values} \\
\hline
\texttt{p} & Output format & \texttt{lmp} \\
\texttt{c} & Crystal structure & \texttt{sc}, \texttt{bcc}, \texttt{fcc} \\
\texttt{m} & Magnetic order & \texttt{fm}, \texttt{afm} \\
$r_{\mathrm{NN}}$ & First-nearest-neighbour distance at the equilibrium volume of the non-magnetic reference system & \AA \\
$B_0$ & Bulk modulus of the non-magnetic reference system & GPa \\
$T_{\mathrm{ord}}$ & Curie temperature ($T_{\mathrm C}$) for FM order or Néel temperature ($T_{\mathrm N}$) for AFM order & K \\
$\omega_s$ & Spontaneous volume magnetostriction at zero temperature & dimensionless \\
$K_1$ & First magnetocrystalline-anisotropy constant at zero temperature & J\,m$^{-3}$ \\
$\eta$ & Relative pressure derivative of $K_1$, defined as $K_1^{-1}\,\partial K_1/\partial P$ & GPa$^{-1}$ \\
$b_1$ & First anisotropic magnetoelastic constant & MPa \\
$b_2$ & Second anisotropic magnetoelastic constant & MPa \\
\hline
\end{tabularx}
\end{table}

\begin{figure}[htbp]
\centering
\includegraphics[width=0.78\linewidth]{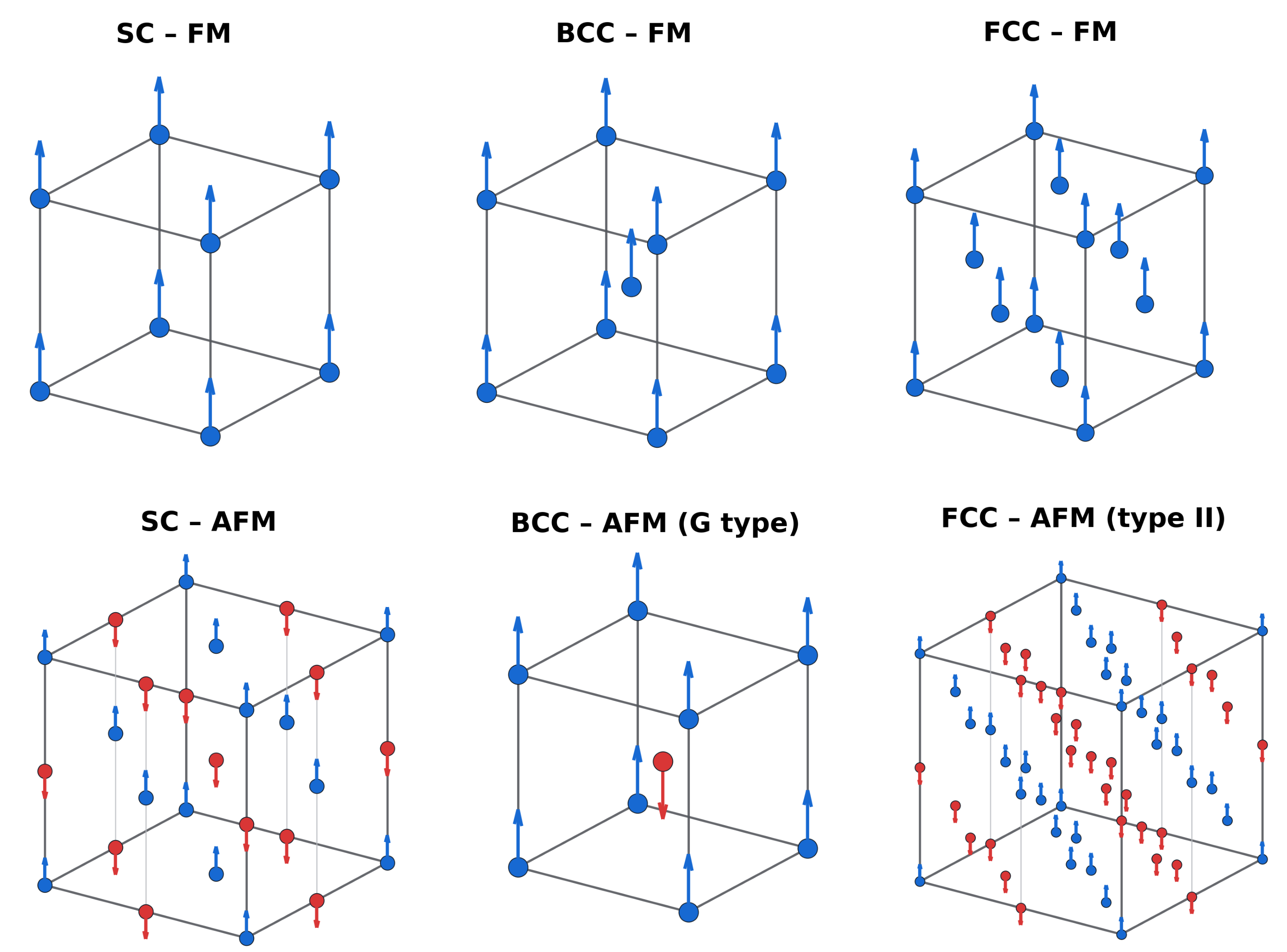}
\caption{Unit cells of the crystal structures and magnetic configurations implemented in \textsc{MagPot} v1.0. }
\label{fig:crystals}
\end{figure}

The distance $r_{\mathrm{NN}}$ in the input file is always the first-neighbour distance of the underlying magnetic cubic sublattice. The distance at which the magnetic functions are evaluated is denoted by $R_0$ in Section \ref{subsec:hamiltonian}. For SC, BCC and FCC ferromagnets, and for SC and BCC antiferromagnets, $R_0=r_{\mathrm{NN}}$. For the FCC type-II antiferromagnetic model, only second-neighbour interactions are retained and therefore $R_0=a_0=\sqrt{2}\,r_{\mathrm{NN}}$, where $a_0$ is the conventional cubic lattice parameter \cite{korniienko_NiO_2026}. This distinction is handled internally by the program. Table~\ref{tab:implemented_models} lists the supported systems in version~1.0 and the neighbour shell used to parameterise the magnetic interactions. The corresponding unit cells of these systems are shown in Fig.\ref{fig:crystals}.

\begin{table}[htbp]
\centering
\caption{Crystal structures and collinear magnetic configurations implemented in \textsc{MagPot} v1.0.}
\label{tab:implemented_models}
\small
\setlength{\tabcolsep}{3.5pt}
\renewcommand{\arraystretch}{1.18}
\begin{tabularx}{\textwidth}{@{}>{\centering\arraybackslash}p{1.3cm}>{\centering\arraybackslash}p{1.25cm}>{\raggedright\arraybackslash}p{1.8cm}>{\centering\arraybackslash}p{2.2cm}X@{}}
\hline
\textbf{Lattice} & \textbf{Order} & \textbf{Shell} & \textbf{$R_0$} & \textbf{Assumption} \\
\hline
SC & FM & 6 first neighbours & $r_{\mathrm{NN}}$ & All first-neighbour moments are parallel. \\
SC & AFM & 6 first neighbours & $r_{\mathrm{NN}}$ & All first-neighbour moments are antiparallel. \\
BCC & FM & 8 first neighbours & $r_{\mathrm{NN}}$ & All first-neighbour moments are parallel. \\
BCC & AFM (type G) & 8 first neighbours & $r_{\mathrm{NN}}$ & All eight first-neighbour moments are antiparallel. \\
FCC & FM & 12 first neighbours & $r_{\mathrm{NN}}$ & All first-neighbour moments are parallel. \\
FCC & AFM (type II) & 6 second neighbours & $a_0=\sqrt{2}\,r_{\mathrm{NN}}$ & The 12 first neighbours comprise six parallel and six antiparallel pairs and are neglected. The six second neighbours are antiparallel and form the nearest-neighbour shell of each of the four interpenetrating SC networks. \\
\hline
\end{tabularx}
\end{table}

The FCC-AFM type-II configuration requires special treatment because a collinear FCC lattice cannot have all twelve first neighbours antiparallel due to geometrical frustration. In this configuration, the six second neighbours are antiparallel and form the nearest-neighbour shell of each of the four interpenetrating simple-cubic (SC) networks. Following the approach adopted for NiO \cite{korniienko_NiO_2026}, version~1.0 retains these second-neighbour interactions while neglecting both parallel and antiparallel first-neighbour interactions and interactions between different SC networks. Accordingly, the SC-AFM expressions are applied independently to each network. In LAMMPS, four atom types are used to identify the four SC networks, and non-zero magnetic coefficients are assigned only to pairs belonging to the same network; interactions between different networks are set to zero. The atom types therefore identify the crystallographic networks rather than the spin directions, which are specified independently by the spin vectors. In the NiO-like $2\times2\times2$ magnetic cell, each network contains two magnetic sublattices, giving eight magnetic sublattices in total. The user must assign the atom types in the LAMMPS structure consistently with this convention. The conventional FCC cell contains four magnetic atoms.

After reading the input file, \textsc{MagPot} prints the calculated Bethe--Slater parameters and the corresponding LAMMPS \texttt{pair\_coeff} commands. The output can be redirected to a file, for example,
\begin{verbatim}
python3 magpot.py > results.dat
\end{verbatim}
For the exchange interaction, the generated command has the form
\begin{lstlisting}[
    basicstyle=\ttfamily\small,
    breaklines=true,
    breakatwhitespace=true,
    columns=fullflexible
]
pair_coeff * * spin/exchange exchange Rc alphaJ gammaJ deltaJ offset no
\end{lstlisting}
whereas the Neel contribution is written as
\begin{lstlisting}[
    basicstyle=\ttfamily\small,
    breaklines=true,
    breakatwhitespace=true,
    columns=fullflexible
]
pair_coeff * * spin/neel neel Rc alphal gammal deltal alphaq gammaq deltaq
\end{lstlisting}
The parameters $\alpha_\Lambda$, $\gamma_\Lambda$ and $\delta_\Lambda$ correspond to the Bethe--Slater functions $\Lambda=J,l,q$, respectively. The cutoff $R_c$ is printed in \AA; the energy parameters are expressed in eV when used with the LAMMPS \texttt{metal} unit convention, and the dimensionless shape parameters are unitless. The cutoff must also be specified in the corresponding \texttt{pair\_style} declaration. A typical magnetic pair-style definition is
\begin{verbatim}
pair_style hybrid/overlay eam/alloy spin/exchange Rc spin/neel Rc
\end{verbatim}
with an appropriate EAM/MEAM coefficient supplied separately. Simulations using the SPIN pair styles require a spin-capable atom style (\texttt{atom\_style spin}).

\subsection{Magnetic Hamiltonian and Bethe--Slater parametrisation}
\label{subsec:hamiltonian}

The total potential energy of the spin--lattice model is decomposed into a spin-independent reference contribution and a magnetic contribution:
\begin{equation}
U_{\mathrm{pot}} =
U_{\mathrm{ref}}(\{\mathbf r_i\})+
H_{\mathrm{mag}}(\{\mathbf r_i,\mathbf s_i\}),
\end{equation}
where $U_{\mathrm{ref}}$ describes the underlying atomic interactions and is provided by an independently selected interatomic potential, while $H_{\mathrm{mag}}$ contains the magnetic interactions generated by \textsc{MagPot}. This additive decomposition makes the approach modular, allowing the magnetic contribution to be combined with different compatible interatomic potentials through the \texttt{hybrid/overlay} functionality of LAMMPS. In the present implementation, the magnetic contribution consists of a Heisenberg exchange term and a two-ion Néel interaction:
\begin{equation}
H_{\mathrm{mag}}=H_{\mathrm{ex}}+H_{\mathrm{N}}.
\end{equation}
The exchange contribution is
\begin{equation}
H_{\mathrm{ex}}=-\frac{1}{2}\sum_{i\ne j}J(r_{ij})\,\mathbf{s}_i\cdot\mathbf{s}_j,
\label{eq:exchange_hamiltonian}
\end{equation}
where $r_{ij}=|\mathbf r_i-\mathbf r_j|$ and $\mathbf{s}_i$ is the unit vector describing the direction of the magnetic moment at site $i$. With this sign convention, positive $J$ favours parallel alignment, whereas negative $J$ favours antiparallel alignment.

The Néel interaction introduces the dependence of the magnetic energy on the orientation of the spin moments relative to the interatomic bond direction. Defining $\mathbf r_{ij}=\mathbf r_i-\mathbf r_j$,
$r_{ij}=|\mathbf r_{ij}|$ and $\mathbf e_{ij}=\mathbf r_{ij}/r_{ij}$, the two-ion Néel
Hamiltonian can be written as \cite{lacheisserie_magnetostriction_1993}
\begin{align}
H_{\mathrm N}=-\frac{1}{2}\sum_{i\ne j}\Bigg\{&
l_1(r_{ij})\left[
(\mathbf e_{ij}\cdot\mathbf s_i)
(\mathbf e_{ij}\cdot\mathbf s_j)-\frac{\mathbf s_i\cdot\mathbf s_j}{3}
\right]
\nonumber\\
&+q_1(r_{ij})\left[
(\mathbf e_{ij}\cdot\mathbf s_i)^2-\frac{\mathbf s_i\cdot\mathbf s_j}{3}
\right]
\left[
(\mathbf e_{ij}\cdot\mathbf s_j)^2-\frac{\mathbf s_i\cdot\mathbf s_j}{3}
\right]
\nonumber\\
&+q_2(r_{ij})\left[
(\mathbf e_{ij}\cdot\mathbf s_i)
(\mathbf e_{ij}\cdot\mathbf s_j)^3
+
(\mathbf e_{ij}\cdot\mathbf s_j)
(\mathbf e_{ij}\cdot\mathbf s_i)^3
\right]\Bigg\},
\label{eq:neel_hamiltonian}
\end{align}
where the radial functions $l_1$, $q_1$ and $q_2$ are related to
the dipolar and quadrupolar functions $l$ and $q$ through
\begin{equation}
l_1(r)=l(r)+\frac{12}{35}q(r),\qquad
q_1(r)=\frac{9}{5}q(r),\qquad
q_2(r)=-\frac{2}{5}q(r).
\label{eq:neel_coefficients}
\end{equation}
For collinear spin configurations, the Néel energy reduces to
different angular expressions depending on whether the pair is
parallel or antiparallel. For parallel pairs
($\mathbf{s}_i\cdot\mathbf{s}_j=1$),
\begin{equation}
H_{\mathrm N}^{\uparrow\uparrow}
=-\frac{1}{2}\sum_{i\ne j}\left\{
l(r_{ij})\left[
(\mathbf e_{ij}\cdot\mathbf s_i)^2-\frac{1}{3}
\right]
+
q(r_{ij})\left[
(\mathbf e_{ij}\cdot\mathbf s_i)^4
-\frac{6}{7}(\mathbf e_{ij}\cdot\mathbf s_i)^2
+\frac{3}{35}
\right]\right\},
\label{eq:neel_fm}
\end{equation}
whereas for antiparallel pairs
($\mathbf{s}_i\cdot\mathbf{s}_j=-1$),
\begin{equation}
H_{\mathrm N}^{\uparrow\downarrow}
=-\frac{1}{2}\sum_{i\ne j}\left\{
-l(r_{ij})\left[
(\mathbf e_{ij}\cdot\mathbf s_i)^2-\frac{1}{3}
\right]
+
q(r_{ij})\left[
\frac{13}{5}(\mathbf e_{ij}\cdot\mathbf s_i)^4
+\frac{6}{7}(\mathbf e_{ij}\cdot\mathbf s_i)^2
+\frac{11}{35}
\right]\right\}.
\label{eq:neel_afm}
\end{equation}
The dipolar function $l(r)$ determines the leading anisotropic magnetoelastic contribution, while the quadrupolar function $q(r)$ controls the cubic magnetocrystalline anisotropy. The relations used to connect $l(r_0)$ and its derivative to $b_1$ and $b_2$ neglect the quadrupolar contribution to the magnetoelastic energy, following the approximation adopted in the cubic spin--lattice parametrisation of Nieves \textit{et al.} \cite{nieves_spin-lattice_2021,chikazumi_physics_2009,ohandley_modern_1999}. The function $q(r)$ is then determined independently from $K_1$ and its pressure dependence.

The distance dependence of the three magnetic functions is represented by the Bethe--Slater form used in the LAMMPS SPIN package \cite{tranchida_massively_2018}:
\begin{equation}
\Lambda(r)=4\alpha_\Lambda\left(\frac{r}{\delta_\Lambda}\right)^2
\left[1-\gamma_\Lambda\left(\frac{r}{\delta_\Lambda}\right)^2\right]
\exp\!\left[-\left(\frac{r}{\delta_\Lambda}\right)^2\right]
\Theta(R_{c,\Lambda}-r),
\label{eq:bethe_slater}
\end{equation}
where $\Lambda=J,l,q$, $\Theta$ is the Heaviside step function and $R_{c,\Lambda}$ is the cutoff radius. For each function, the analytical parametrisation requires its value and first radial derivative at the reference interaction distance $R_0$. We define
\begin{equation}
D_\Lambda=R_0\left.\frac{\partial\Lambda}{\partial r}\right|_{r=R_0},
\qquad \Lambda_0=\Lambda(R_0).
\label{eq:radial_derivative_definition}
\end{equation}
The three Bethe--Slater parameters are then \cite{nieves_spin-lattice_2021}
\begin{equation}
\delta_\Lambda=R_0,\qquad
\alpha_\Lambda=\frac{e}{8}\left(2\Lambda_0-D_\Lambda\right),\qquad
\gamma_\Lambda=\frac{D_\Lambda}{D_\Lambda-2\Lambda_0}.
\label{eq:bethe_slater_parameters}
\end{equation}
The cutoff is selected halfway between the interaction shell at $R_0$ and the next shell of the corresponding interaction network, so that no additional neighbour shell is included. In version~1.0, the same geometrical cutoff is used for $J(r)$, $l(r)$ and $q(r)$ for each selected lattice/order combination.

The values of $J_0$ and $D_J$ are obtained by relating the exchange energy to the Curie or Néel temperature and the spontaneous volume magnetostriction. The values of $l_0$ and $D_l$ follow from the cubic magnetoelastic energy and the constants $b_1$ and $b_2$. Finally, $q_0$ and $D_q$ are determined by matching the Néel energy to the first cubic magnetocrystalline-anisotropy constant and its pressure dependence. The following notation is used in the expressions implemented in the code:
\begin{equation}
\Omega=\frac{V_0}{n},\qquad B=10^9B_0\;\mathrm{Pa},\qquad
\eta=\frac{1}{K_1}\frac{\partial K_1}{\partial P},
\label{eq:volume_modulus_eta}
\end{equation}
where $V_0=a_0^3$ is the conventional cubic-cell volume and $n$ is the number of atoms in that cell ($n=1$, 2 and 4 for SC, BCC and FCC, respectively). The sign of $\omega_s$ must follow the convention used to obtain the material input; the signs in Table~\ref{tab:magnetic_formulas} are those implemented for the FM and AFM definitions adopted in Refs.~\cite{nieves_spin-lattice_2021,korniienko_NiO_2026}. For a cubic crystal, $B=(C_{11}+2C_{12})/3$, so the combination of elastic constants $C_{11}+2C_{12}$ appearing in the exchange-magnetostriction expressions can be replaced by $3B$. This is why \textsc{MagPot} requires only the bulk modulus, rather than the individual elastic constants $C_{11}$ and $C_{12}$. In the derivative of $q$, the numerical value of $B_0$ is in GPa and that of $\eta$ is in GPa$^{-1}$. Internally, distances are converted from \AA{} to metres, $B_0$ from GPa to Pa, and $b_1,b_2$ from MPa to Pa. Energy-valued quantities are finally converted from joules to eV using $1~\mathrm{J}=6.241509\times10^{18}~\mathrm{eV}$.

\begin{table}[htbp]
\centering
\caption{Reference values and scaled radial derivatives used by \textsc{MagPot} version~1.0. The derivative is defined as $D_\Lambda=R_0\Lambda'(R_0)$. In the quadrupolar rows, $A_K=1-B_0\eta$, with $B_0$ in GPa and $\eta$ in GPa$^{-1}$.}
\label{tab:magnetic_formulas}
\scriptsize
\setlength{\tabcolsep}{2pt}
\renewcommand{\arraystretch}{1.45}
\begin{tabularx}{\textwidth}{@{}>{\raggedright\arraybackslash}p{1.7cm}>{\centering\arraybackslash}p{0.8cm}>{\centering\arraybackslash}X>{\centering\arraybackslash}X@{}}
\hline
\textbf{Model} & \textbf{Term} & $\boldsymbol{\Lambda_0}$ & $\boldsymbol{D_\Lambda}$ \\
\hline
SC--FM & $J$ & $\tfrac{k_{\mathrm B}T_{\mathrm C}}{2}$ & $+\omega_s B\Omega$ \\
SC--FM & $l$ & $-\tfrac{1}{2}\Omega b_2$ & $-\Omega b_1$ \\
SC--FM & $q$ & $+\tfrac{1}{2}\Omega K_1$ & $3q_0 A_K$ \\
\hline
SC--AFM & $J$ & $-\tfrac{k_{\mathrm B}T_{\mathrm N}}{2}$ & $-\omega_s B\Omega$ \\
SC--AFM & $l$ & $+\tfrac{1}{2}\Omega b_2$ & $+\Omega b_1$ \\
SC--AFM & $q$ & $+\tfrac{5}{26}\Omega K_1$ & $3q_0 A_K$ \\
\hline
BCC--FM & $J$ & $+\tfrac{3k_{\mathrm B}T_{\mathrm C}}{8}$ & $+\tfrac{3}{4}\omega_s B\Omega$ \\
BCC--FM & $l$ & $-\tfrac{3}{8}\Omega b_1$ & $+\tfrac{3}{8}\Omega(b_1-3b_2)$ \\
BCC--FM & $q$ & $-\tfrac{9}{16}\Omega K_1$ & $3q_0 A_K$ \\
\hline
BCC--AFM (G) & $J$ & $-\tfrac{3k_{\mathrm B}T_{\mathrm N}}{8}$ & $-\tfrac{3}{4}\omega_s B\Omega$ \\
BCC--AFM (G) & $l$ & $+\tfrac{3}{8}\Omega b_1$ & $-\tfrac{3}{8}\Omega(b_1-3b_2)$ \\
BCC--AFM (G) & $q$ & $-\tfrac{45}{208}\Omega K_1$ & $3q_0 A_K$ \\
\hline
FCC--FM & $J$ & $+\tfrac{k_{\mathrm B}T_{\mathrm C}}{4}$ & $+\tfrac{1}{2}\omega_s B\Omega$ \\
FCC--FM & $l$ & $+\tfrac{1}{2}\Omega\left(\tfrac{b_2}{2}-b_1\right)$ & $+\Omega\left(b_1-\tfrac{3b_2}{2}\right)$ \\
FCC--FM & $q$ & $-\Omega K_1$ & $3q_0 A_K$ \\
\hline
FCC--AFM (II) & $J$ & $-\tfrac{k_{\mathrm B}T_{\mathrm N}}{2}$ & $-\omega_s B\Omega$ \\
FCC--AFM (II) & $l$ & $+\tfrac{1}{2}\Omega b_2$ & $+\Omega b_1$ \\
FCC--AFM (II) & $q$ & $+\tfrac{5}{26}\Omega K_1$ & $3q_0 A_K$ \\
\hline
\end{tabularx}
\end{table}

Table~\ref{tab:magnetic_formulas} collects the values and radial derivatives implemented in version~1.0. For the BCC type-G antiferromagnet, all eight first neighbours are antiparallel, see Fig.\ref{fig:crystals}. The quadrupolar prefactor in Table~\ref{tab:magnetic_formulas} is obtained by summing the AFM angular expression in Eq.~(\ref{eq:neel_afm}) over the eight BCC bond directions and matching the energy difference between the $[110]$ and $[100]$ spin orientations to the cubic anisotropy energy $V_0K_1/4$. For FCC type-II AFM, the SC-AFM expressions are used for each of the four interpenetrating SC networks; the cell volume and atom count correspond to the conventional FCC cell ($n=4$), and $R_0$ is the FCC second-neighbour distance. These assumptions are part of the model definition and should be retained when constructing the LAMMPS structure and atom types.

\subsection{Parameterisation workflow}
\label{subsec:workflow}

The workflow used to generate a magnetic potential with \textsc{MagPot} is divided into two stages. First, the equilibrium geometry and bulk modulus are determined for the non-magnetic reference system. Second, the magnetic and magnetoelastic material properties are combined with these structural quantities in \texttt{input.dat} to obtain the Bethe--Slater parameters.

\paragraph{Non-magnetic equation of state.}
An interatomic potential describing the non-magnetic lattice contribution (for example, EAM, MEAM, Lennard--Jones,  Buckingham, SNAP, etc.) is selected first. The magnetic exchange and Néel interactions are disabled, and the total energy is calculated for a set of volumes around the equilibrium volume. The resulting energy--volume data are fitted to an equation of state (EOS), such as the Vinet \cite{vinet_temperature_1987} or Murnaghan \cite{murnaghan_compressibility_1944} form, to obtain the equilibrium conventional-cell volume $V_0$ and bulk modulus $B_0$. The equilibrium lattice parameter is then $a_0=V_0^{1/3}$ for the conventional cubic cell, and the first-neighbour distance supplied to \textsc{MagPot} is obtained from
\begin{equation}
r_{\mathrm{NN}}=\begin{cases}
a_0, & \mathrm{SC},\\[2pt]
\dfrac{\sqrt{3}}{2}a_0, & \mathrm{BCC},\\[5pt]
\dfrac{a_0}{\sqrt{2}}, & \mathrm{FCC}.
\end{cases}
\label{eq:nearest_neighbour_distances}
\end{equation}
The FCC-AFM type-II branch subsequently converts this input first-neighbour distance to the second-neighbour interaction distance $R_0=a_0$. In the distributed BCC-Fe example, the non-magnetic EOS is fitted to the energy--volume data generated with LAMMPS; the example script uses a Vinet EOS fit to obtain $B_0$ and $V_0$, see Section \ref{subsec:test_eos}.

\paragraph{Magnetic and magnetoelastic input properties.}
The remaining material properties can be obtained from experiment or from first-principles calculations. The transition temperature $T_{\mathrm C}$ or $T_{\mathrm N}$ and the spontaneous volume magnetostriction $\omega_s$ may be taken from experimental measurements or suitable theoretical calculations \cite{Nieves_FeRh_2026}. The magnetocrystalline anisotropy $K_1$ can be evaluated from spin--orbit-coupled DFT total energies for different magnetisation directions. Its pressure derivative $\eta$ is obtained by repeating the calculation at several volumes or pressures and fitting the resulting $K_1(P)$ dependence. The anisotropic magnetoelastic constants $b_1$ and $b_2$ can be obtained from strain-dependent energy calculations; the corresponding DFT workflows are available in \textsc{MAELAS} and can be carried out using VASP as the electronic-structure engine \cite{nieves_maelas_2021,nieves_maelas_2022,nieves_automated_2023}. Experimental values may be used instead whenever available and consistent with the definitions adopted by the model.

Once these quantities have been collected and converted to the units specified in Table~\ref{tab:input_parameters}, they are written to \texttt{input.dat}. Running \texttt{magpot.py} produces the Bethe--Slater parameters and LAMMPS coefficient commands. These magnetic terms are then combined with the non-magnetic interatomic potential in a spin--lattice simulation. The separation between the non-magnetic EOS calculation and the subsequent magnetic parameterisation is intentional: the input values $r_{\mathrm{NN}}$ and $B_0$ describe the reference lattice potential without magnetic interactions, while $T_{\mathrm{ord}}$, $\omega_s$, $K_1$, $\eta$, $b_1$ and $b_2$ determine the magnetic part of the model.

\section{Examples and tests}
\label{sec:examples}

The examples distributed with \textsc{MagPot} version~1.0 are based on body-centred cubic (BCC) iron. They illustrate the complete parameterisation workflow and test whether the generated spin--lattice potential reproduces the material properties used as input. In this test, the non-magnetic interactions were described using the EAM potential for Fe developed by Chamati et al. \cite{Chamati2006}, whereas the magnetic contribution is given by the exchange and N\'eel terms generated by \textsc{MagPot}. The calculations are carried out with LAMMPS and its SPIN package \cite{thompson_lammps_2022,tranchida_massively_2018}. In the first step, the EOS of the spin-independent reference potential is analysed to determine the required input parameters, \(r_{\mathrm{NN}}\) and \(B_0\). These structural quantities are then combined with the target magnetic and magnetoelastic properties to generate the magnetic contribution to the spin–lattice potential using \textsc{MagPot}. In the next step, we assess whether the resulting spin–lattice model reproduces the target properties, including the Curie temperature, spontaneous volume magnetostriction, pressure dependence of the magnetocrystalline anisotropy, and anisotropic magnetoelastic constants. Similar tests for the other supported symmetries are discussed in Section \ref{subsec:tests_sym}.

\subsection{Step 1: Non-magnetic equation of state}
\label{subsec:test_eos}

The first calculation establishes the equilibrium structural parameters required by \textsc{MagPot}. A set of seven BCC Fe conventional-cell volumes is evaluated using the EAM/alloy potential without the magnetic exchange and N\'eel interactions (which have not yet been determined at this stage). The calculated energy--volume data are fitted to a Vinet equation of state. The fit gives an equilibrium conventional-cell volume of $V_0=23.5534$~\AA$^3$ and a bulk modulus of $B_0=178.762$~GPa. Since the conventional BCC cell contains two atoms, the corresponding lattice parameter is $a_0=V_0^{1/3}=2.86650$~\AA, and the nearest-neighbour distance is $r_{\mathrm{NN}}=\sqrt{3}a_0/2=2.48246$~\AA. These values are the structural inputs used in the BCC Fe \texttt{input.dat} file.

\begin{figure}[htbp]
\centering
\includegraphics[width=\linewidth]{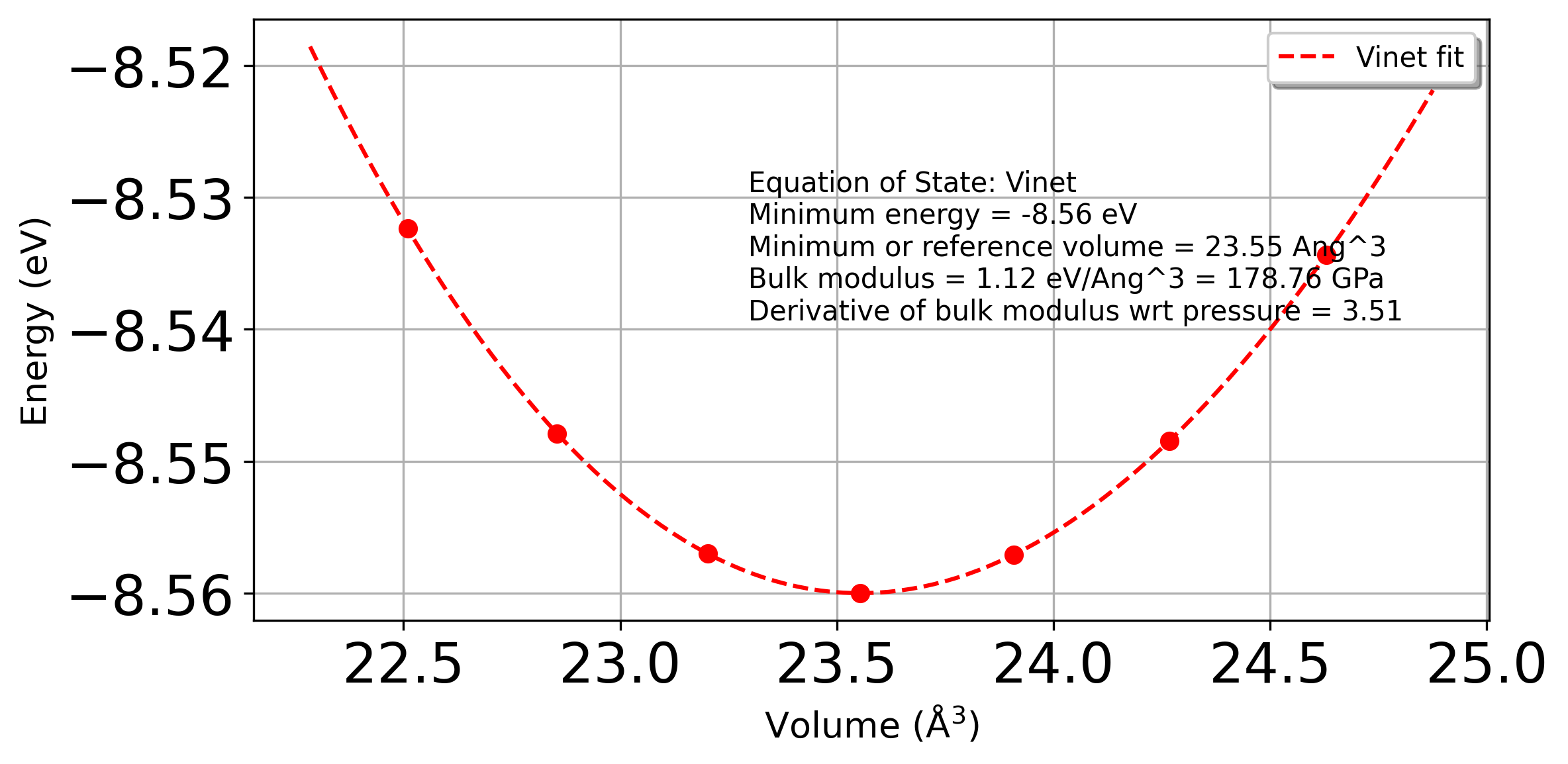}
\caption{Energy--volume data for non-magnetic BCC Fe and the Vinet EOS fit used to determine the equilibrium volume and bulk modulus.}
\label{fig:fe_eos}
\end{figure}

\subsection{Step 2: input data and generated Bethe--Slater parameters}
\label{subsec:fe_parameters}

Next, we fill in the \texttt{input.dat} file for the ferromagnetic BCC Fe example. The first-neighbour distance ($r_{\mathrm{NN}}=2.48246$~\AA) and bulk modulus ($B_0=178.762$~GPa) were obtained in the preliminary non-magnetic EOS calculation (Section \ref{subsec:test_eos}). The remaining entries ($T_C$, $\omega_s$, $K_1(P=0)$, $K_1^{-1}\,\partial K_1/\partial P$, $b_1$ and $b_2$) are the target magnetic and magnetoelastic properties used to generate the potential, which are taken from literature. The input data are listed in Table~\ref{tab:fe_input_model}. The corresponding \texttt{input.dat} file and the LAMMPS input files are included in the \texttt{Examples/} directory of the code distribution.

\begin{table}[htbp]
\centering
\caption{BCC Fe parameters supplied in the \texttt{input.dat} file of \textsc{MagPot} v1.0 example. The data for $r_{\mathrm{NN}}$ and $B_0$ are obtained from the EOS analysis using the non-magnetic potential (Section \ref{subsec:test_eos}), while the other data are taken from literature.}
\label{tab:fe_input_model}
\small
\renewcommand{\arraystretch}{1.15}
\begin{tabular}{p{3.0cm}p{2.7cm}p{1.8cm}}
\hline
\textbf{Parameter} & \textbf{Input value} & \textbf{Unit} \\
\hline
$r_{\mathrm{NN}}$ & 2.482458435 & \AA \\
$B_0$ & 178.762 & GPa \\
$T_{\mathrm C}$ & 1043.0 \cite{ohandley_modern_1999} & K \\
$\omega_s$ & $1.16\times10^{-2}$ \cite{shimizu_forced_1978} & -- \\
$K_1(P=0)$ & 55\,000 \cite{getzlaff_fundamentals_2008} & J\,m$^{-3}$ \\
$K_1^{-1}\,\partial K_1/\partial P$ & $-0.073$ \cite{sawaoka_effect_1975} & GPa$^{-1}$ \\
$b_1$ & $-3.3$ \cite{ohandley_modern_1999} & MPa \\
$b_2$ & 10.5 \cite{ohandley_modern_1999} & MPa \\
\hline
\end{tabular}
\end{table}

Using these inputs, \textsc{MagPot} generates the Bethe--Slater parameters collected in Table~\ref{tab:fe_bethe_slater}. The three functions share the same cutoff and reference distance in this example. The values of $\alpha_\Lambda$ are given in eV, as required by the LAMMPS \texttt{metal} unit system; $\gamma_\Lambda$ is dimensionless and $\delta_\Lambda$ and $R_{c,\Lambda}$ are given in \AA. The resulting exchange and N\'eel coefficients are used together with the separately specified EAM/alloy coefficient through the LAMMPS \texttt{hybrid/overlay} pair style. Table~\ref{tab:fe_model_comparison} compares the target values entered in \texttt{input.dat} with the properties recovered from the corresponding LAMMPS tests, finding good agreement for all quantities.

\begin{table}[htbp]
\centering
\caption{Bethe--Slater parameters generated by \textsc{MagPot} version~1.0 for BCC ferromagnetic Fe. The cutoff $R_c$ is common to the three magnetic functions in this example.}
\label{tab:fe_bethe_slater}
\small
\renewcommand{\arraystretch}{1.15}
\begin{tabular}{lrrrr}
\hline
\textbf{Function} & $\boldsymbol{\alpha_\Lambda}$ (eV) & $\boldsymbol{\gamma_\Lambda}$ & $\boldsymbol{\delta_\Lambda}$ (\AA) & $\boldsymbol{R_c}$ (\AA) \\
\hline
$J(r)$ & $-1.5938382\times10^{-2}$ & 2.4370721 & 2.4824584 & 2.6744773 \\
$l(r)$ & $ 3.8774818\times10^{-4}$ & 0.8405797 & 2.4824584 & 2.6744773 \\
$q(r)$ & $ 3.1022500\times10^{-5}$ & 1.0498146 & 2.4824584 & 2.6744773 \\
\hline
\end{tabular}
\end{table}

\begin{table}[htbp]
\centering
\caption{Comparison between the target material parameters supplied to \textsc{MagPot} and the values recovered from the LAMMPS validation tests for BCC Fe.}
\label{tab:fe_model_comparison}
\small
\setlength{\tabcolsep}{4pt}
\renewcommand{\arraystretch}{1.18}
\begin{tabularx}{\textwidth}{@{}X>{\centering\arraybackslash}p{2.4cm}>{\centering\arraybackslash}p{2.6cm}>{\centering\arraybackslash}p{1.5cm}@{}}
\hline
\textbf{Property} & \textbf{Input target} & \textbf{Model result} & \textbf{Unit} \\
\hline
$T_{\mathrm C}$ & 1043 & 900 & K \\
$\omega_s$ & 0.01160 & 0.011441 & -- \\
$K_1(P=0)$ & 55.000 & 55.070 & kJ\,m$^{-3}$ \\
$K_1^{-1}\,\partial K_1/\partial P$ & $-0.07300$ & $-0.073038$ & GPa$^{-1}$ \\
$b_1$ & $-3.30$ & $-3.720$ & MPa \\
$b_2$ & 10.50 & 10.283 & MPa \\
\hline
\end{tabularx}
\end{table}

\subsection{Curie-temperature test}
\label{subsec:test_tc}

\begin{figure}[htbp]
\centering
\includegraphics[width=0.72\linewidth]{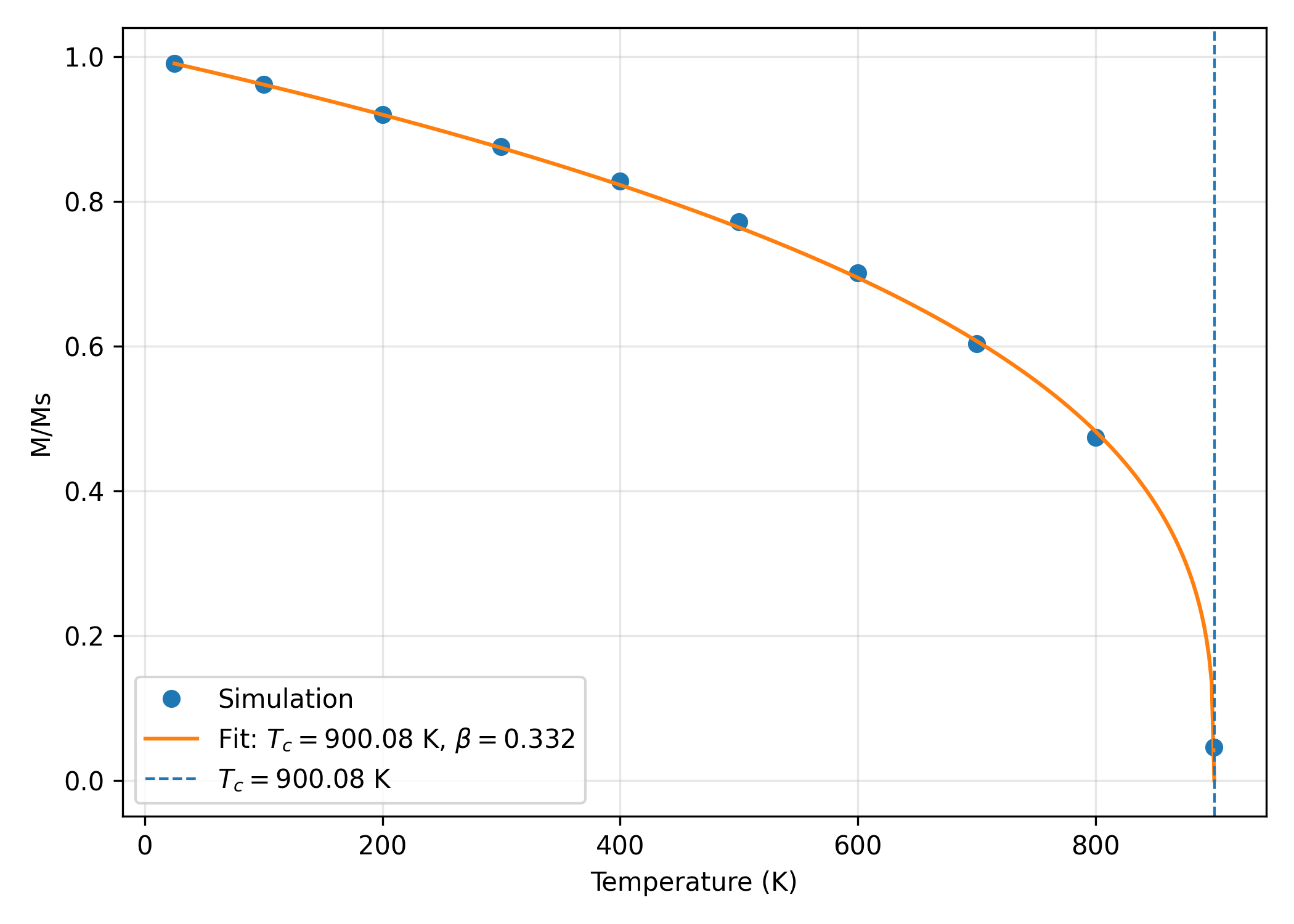}
\caption{Normalised magnetisation of BCC Fe as a function of temperature and fit to the critical law in Eq.~(\ref{eq:tc_critical_fit}).}
\label{fig:fe_tc}
\end{figure}

The Curie temperature is tested by applying a Langevin thermostat \cite{tranchida_massively_2018} on the spin system at a sequence of temperatures and monitoring the normalised magnetisation. The resulting $M(T)/M_s$ data are fitted to the critical-law expression
\begin{equation}
\frac{M(T)}{M_s}=\left(1-\frac{T}{T_{\mathrm C}}\right)^\beta,
\qquad T<T_{\mathrm C},
\label{eq:tc_critical_fit}
\end{equation}
where both $T_{\mathrm C}$ and the exponent $\beta$ are fitting parameters, as shown in Fig.\ref{fig:fe_tc}. The fit reported in the example gives $T_{\mathrm C}=900$~K and $\beta=0.332$. The target Curie temperature used to generate the potential is 1043~K. This deviation is reasonable within the Mean-Field Approximation (MFA) used to compute $J(r)$. We note that the agreement between simulated and experimental temperature-dependent magnetization curves can be further improved by applying a temperature-rescaling scheme, in which the physical temperature is mapped onto an effective simulation temperature that controls the thermal fluctuations of the spin system \cite{evans_quantitative_2015}.

\subsection{Spontaneous volume magnetostriction}
\label{subsec:test_ws}

\begin{figure}[htbp]
\centering
\includegraphics[width=0.78\linewidth]{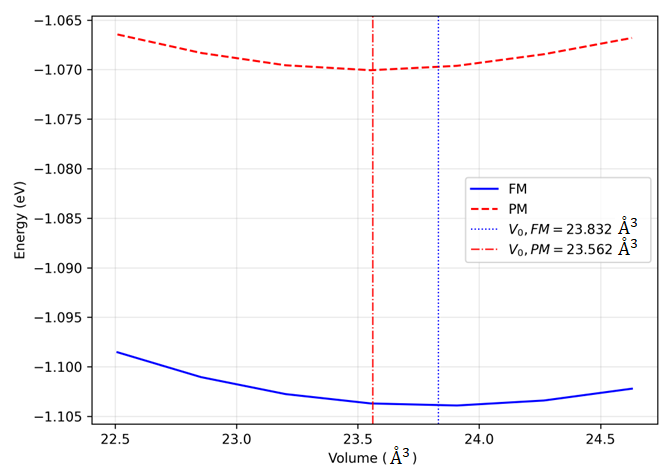}
\caption{Comparison of the ferromagnetic and paramagnetic energy--volume curves. The vertical lines indicate the equilibrium volumes obtained from the Murnaghan fits. Their difference gives the spontaneous volume magnetostriction according to Eq.~(\ref{eq:ws_test}).}
\label{fig:ws_result}
\end{figure}

The spontaneous volume magnetostriction is evaluated from separate energy--volume curves for the ferromagnetic and paramagnetic configurations. The paramagnetic reference is represented by randomly oriented magnetic moments, while the ferromagnetic calculation uses collinear, parallel moments. The equilibrium volumes are obtained by fitting the corresponding data to the Murnaghan EOS. The fitted equilibrium volumes are
\begin{equation}
V_{\mathrm{FM}}=23.83178~\mathrm{\mathring{A}}^3, \qquad
V_{\mathrm{PM}}=23.56220~\mathrm{\mathring{A}}^3.
\end{equation}
The spontaneous volume magnetostriction is then calculated as
\begin{equation}
\omega_s=\frac{V_{\mathrm{FM}}-V_{\mathrm{PM}}}{V_{\mathrm{PM}}}
=0.0114412.
\label{eq:ws_test}
\end{equation}
This value differs by approximately 1.4\% from the target value $\omega_s=0.0116$ used in the input file. The two EOS and the resulting volume difference is shown in Fig.~\ref{fig:ws_result}.

\subsection{Pressure dependence of $K_1$}
\label{subsec:test_k1}

\begin{figure}[htbp]
\centering
\includegraphics[width=0.72\linewidth]{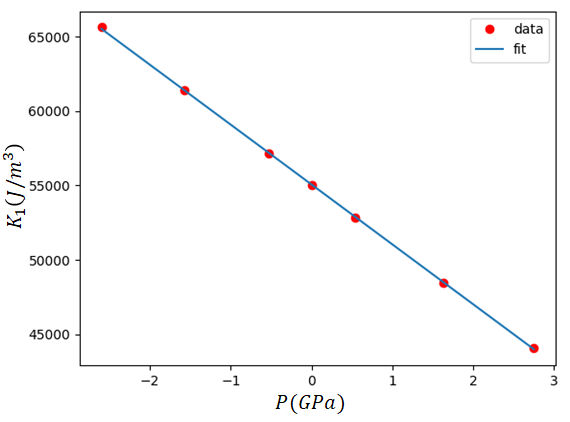}
\caption{Pressure dependence of $K_1$ for BCC Fe. Symbols show the values obtained from the energy differences between magnetisation directions $[110]$ and $[100]$, and the solid line is the linear fit used to determine $K_1(0)$ and $K_1^{-1}\,\partial K_1/\partial P$.}
\label{fig:fe_k1_pressure}
\end{figure}

The pressure dependence of the first magnetocrystalline-anisotropy constant is tested by calculating the energies of BCC Fe with the magnetisation constrained along the crystallographic directions $[100]$ and $[110]$ at a series of volumes. For each volume, the anisotropy constant is obtained from
\begin{equation}
K_1(P)=\frac{4\left[E_{[110]}(P)-E_{[100]}(P)\right]}{V(P)}.
\label{eq:k1_from_energies}
\end{equation}
The seven volume points span approximately $-2.60$ to $2.75$~GPa. A linear fit of the form $K_1(P)=K_1(0)+aP$ yields the following parameters:
\begin{align*}
K_1(0)&=55\,070.0~\mathrm{J\,m^{-3}},\\
a&=-4.0222\times10^{-6}~\mathrm{J\,m^{-3}\,Pa^{-1}}.
\end{align*}
The resulting linear fit is shown in Fig.~\ref{fig:fe_k1_pressure}. The corresponding relative pressure derivative is
\begin{equation}
\left.\frac{1}{K_1}\frac{\partial K_1}{\partial P}\right|_{P=0}
=-0.0730382~\mathrm{GPa}^{-1},
\end{equation}
with a coefficient of determination $R^2=0.999901$. These values agree closely with the input targets of 55\,000~J\,m$^{-3}$ and $-0.073$~GPa$^{-1}$.

\subsection{Anisotropic magnetoelastic constants}
\label{subsec:test_b1b2}

\begin{figure}[htbp]
\centering
\includegraphics[width=\linewidth]{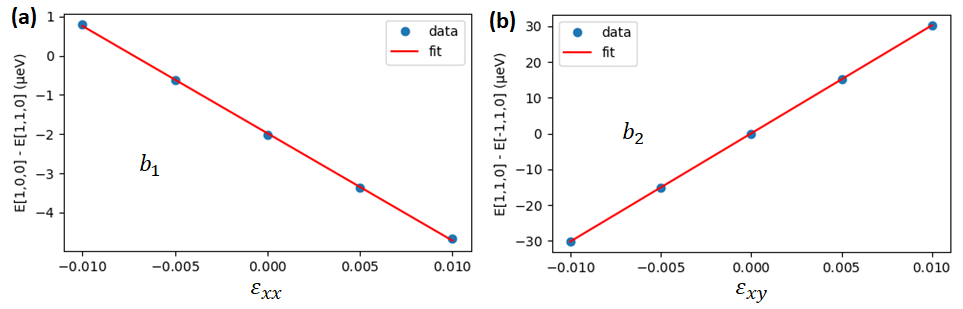}
\caption{Energy differences as a function of strain used to determine the anisotropic magnetoelastic constants of BCC Fe: (a) tetragonal strain $\epsilon_{xx}$ for $b_1$ and (b) shear strain $\epsilon_{xy}$ for $b_2$. Symbols denote the LAMMPS results and solid lines the linear fits performed by MAELAS.}
\label{fig:fe_b1b2}
\end{figure}

The anisotropic magnetoelastic constants are evaluated using the strain--energy procedure implemented in MAELAS mode~2 \cite{nieves_maelas_2022}. The strained atomic configurations generated for this procedure are automatically converted into LAMMPS data files using Atomsk \cite{hirel_atomsk_2015}. The corresponding total energies are then calculated with LAMMPS using the generated spin--lattice potential. For $b_1$, a set of tetragonal strains $\epsilon_{xx}$ is applied and the energy difference between magnetisation directions $[100]$ and $[110]$ is fitted as a linear function of strain. For $b_2$, a set of shear strains $\epsilon_{xy}$ is applied and the corresponding energy difference between directions $[\bar{1}10]$ and $[110]$ is fitted. The values extracted from the LAMMPS energies are $b_1=-3.7199$~MPa and $b_2=10.2828$~MPa, which compare well with the target values in Table~\ref{tab:fe_input_model}. The two energy--strain fits are shown in Fig.~\ref{fig:fe_b1b2}.

\subsection{Tests for the other supported symmetries}
\label{subsec:tests_sym}

We have also verified that \textsc{MagPot} reproduces the cases for FCC-FM like FCC Ni \cite{nieves_spin-lattice_2021}, SC-AFM like FeRh \cite{Nieves_FeRh_2026} and FCC-AFM (type II) like NiO \cite{korniienko_NiO_2026}. The two remaining systems SC-FM and BCC-AFM (type G) have been successfully tested assuming a hypothetical Fe crystal with such a lattice and magnetic order with the target properties given by Table \ref{tab:fe_model_comparison}.

\section{Conclusions and future perspectives}
\label{sec:conclusions}

A Python-based software tool, \textsc{MagPot}, has been presented for the generation of magnetic interaction parameters for spin--lattice molecular dynamics simulations. Starting from a compact set of structural, elastic, magnetic and magnetoelastic properties, the program evaluates the reference values and radial derivatives of the exchange function $J(r)$ and the two Néel functions $l(r)$ and $q(r)$. These quantities are then mapped onto Bethe--Slater functions and written as coefficient commands that can be used with the SPIN package in \textsc{LAMMPS}. In this way, \textsc{MagPot} provides a practical step between the determination of material properties and the construction of an atomistic spin--lattice model.

An important practical advantage of \textsc{MagPot} is its modular architecture, which separates the parameterisation of magnetic interactions from the choice of the underlying interatomic potential. The program generates the magnetic exchange and Néel contributions, which can be added to an independently selected lattice potential using the LAMMPS \texttt{hybrid/overlay} framework. This design makes \textsc{MagPot} suitable for use with a broad range of compatible classical interaction models like EAM, MEAM\cite{nieves_spin-lattice_2021}, Lennard-Jones \cite{Nieves_FeRh_2026}, Buckingham \cite{korniienko_NiO_2026}, reference-free
MEAM \cite{korniienko_NiO_2026}, SNAP \cite{nikolov_temperature_2023}, etc.

Version~1.0 covers SC, BCC and FCC lattices, with FM and selected collinear AFM configurations. The supported cases include the G-type AFM arrangement on BCC, in which the eight first neighbours are antiparallel, and the type-II AFM arrangement on FCC. For the latter, the implementation considers only the six antiparallel second neighbours, which form four interpenetrating simple-cubic networks; interactions between different networks are neglected. The present release generates input in the \textsc{LAMMPS} format and is intended to be used together with a non-magnetic interatomic potential describing the underlying lattice.

All supported systems were carefully examined through separate tests of the non-magnetic equation of state, the Curie temperature, spontaneous volume magnetostriction, the pressure dependence of the first magnetocrystalline-anisotropy constant, and the anisotropic magnetoelastic constants. These tests demonstrate how the material properties used as input can be checked independently against the response of the generated spin--lattice model.

\textsc{MagPot} is designed to complement first-principles tools such as \textsc{MAELAS}. DFT calculations, including workflows based on \textsc{VASP} and \textsc{MAELAS}, can provide magnetocrystalline-anisotropy and magnetoelastic data, while experimental measurements can supply the same quantities when available. The resulting material parameters can be passed to \textsc{MagPot} to construct the distance-dependent magnetic interactions for larger-scale spin--lattice simulations. This parameter-based workflow offers a route for connecting first-principles characterisation with classical atomistic simulations.

Several developments can be considered for future versions. First, the range of supported crystal symmetries could be extended beyond cubic structures to include, for example, tetragonal, hexagonal and orthorhombic systems. Such extensions would require the inclusion of the corresponding symmetry-specific analytical relations and, where necessary, additional magnetoelastic and magnetocrystalline-anisotropy parameters. Second, output generators could be implemented for other molecular-dynamics or atomistic spin--lattice simulation packages, in addition to the current LAMMPS/SPIN format.

In summary, \textsc{MagPot} provides a compact and transparent way to convert material-specific magnetic and magnetoelastic properties into Bethe--Slater parameters suitable for spin--lattice simulations. The current release establishes this workflow for selected cubic systems and provides example calculations for its validation. Extensions to lower-symmetry crystals and additional simulation formats would broaden its applicability to the computational investigation of magnetoelastic materials across length and time scales.

\section*{Funding}
This work was supported by the Ministry of Education, Youth and Sports of the Czech Republic through the e-INFRA CZ (ID:90254) and by project QM4ST (CZ.02.01.01/00/22\_008/0004572). P. N. acknowledges support by grant MU-23-BG22/00168 funded by The Ministry of Universities of Spain; the project MAGCOOLEST, grant PID2025-174502NB-C31, Ministerio de Ciencia, Innovación y Universidades (Spain). D. L. and I. K. acknowledge support from the project No. 25-14529L of the Grant Agency of Czech Republic. R.I. acknowledges funding from the project BETMATFUS, grant PID2023-149089OB-I00, Ministerio de Ciencia, Innovación y Universidades (Spain), the EIT-RM/EU project ExpSkills-REM, grant number UE-22-EXPSKILLS-21104, the Spanish AEI Project NEXPECH-2, grant number MCINN-24-PCI2024-153437, MCINN-23-PID2022-138256NB-C21. P.N. and R.I. would like to acknowledge the Agencia de Ciencia, Competitividad Empresarial e Innovación Asturiana (Sekuens) project MAGNES, SEK-25-GRU-GIC-24-113. Support provided by the EU-MACE and EuMINe COST Actions CA22123 and CA22143, respectively, is gratefully acknowledged as well.

\section*{Data availability}
The datasets generated and analyzed during the current study are available in the GitHub repository \url{https://github.com/pnieves2019/magpot}. The corresponding author, Pablo Nieves (E-mail: \href{mailto:nievespablo@uniovi.es}{nievespablo@uniovi.es}), is the designated individual responsible for receiving and replying to all data sharing requests.

\section*{Code availability}
The source files of the program \textsc{MagPot} are available in the GitHub repository \url{https://github.com/pnieves2019/magpot}.

\section*{Competing interests}
The authors declare no competing interests.

\section*{Author contributions}
P.N. conceived the study. P.N. and I.K. performed the derivation of the theoretical formulas. P.N. and I.K. developed and tested the program and models. S.A., A.F., R.I. and D.L. contributed to the computational modeling analysis and project discussions. All authors contributed to the interpretation of the results, discussed the bcc Fe magnetoelastic properties, reviewed the manuscript, and approved the final version.

\bibliography{sn-bibliography}

\end{document}